\documentclass[aps,pre,twocolumn,superscriptaddress,showkeys]{revtex4-1}
\usepackage{graphicx}

\begin{document}


\title{Modularity allows classification of human brain networks during \\ music and speech perception}

\author{Melia E. Bonomo}
\email{mbonomo@rice.edu.}
\affiliation{Department of Physics and Astronomy, Rice University, Houston, TX, 77005, USA}
\affiliation{Center for Theoretical Biological Physics, Rice University, Houston, TX, 77005, USA}
 
\author{Christof Karmonik}
\affiliation{Center for Performing Arts Medicine, Houston Methodist Hospital, Houston, TX, 77030, USA}
\affiliation{MRI Core, Houston Methodist Research Institute, Houston, TX, 77030, USA}
\affiliation{Department of Radiology, Weill Cornell Medical College, New York, NY, 10065, USA}

\author{Anthony K. Brandt}
\affiliation{Shepherd School of Music, Rice University, Houston, TX, 77005, USA}

\author{J. Todd Frazier}
\affiliation{Center for Performing Arts Medicine, Houston Methodist Hospital, Houston, TX, 77030, USA}


\begin{abstract}
We investigate the use of modularity as a quantifier of whole-brain functional networks. Brain networks are constructed from functional magnetic resonance imaging while subjects listened to auditory pieces that varied in emotivity and cultural familiarity. The results of our analysis reveal high and low modularity groups based on the network configuration during a subject's favorite song, and this classification can predict network reconfiguration during the other auditory pieces. In particular, subjects in the low modularity group show significant brain network reconfiguration during both familiar and unfamiliar pieces. In contrast, the high modularity brain networks appear more robust and only exhibit significant changes during the unfamiliar music and speech.  We also find differences in the stability of module composition for the two groups during each auditory piece. Our results suggest that the modularity of the whole-brain network plays a significant role in the way the network reconfigures during varying auditory processing demands, and it may therefore contribute to individual differences in neuroplasticity capability during therapeutic music engagement.
\end{abstract}


\maketitle

\section{Introduction}

Modular structure is pervasive in biology and plays an important role in optimizing the functional capabilities of different systems \cite{hartwell1999,lorenz2011}.  Broadly speaking, modularity is the degree to which the components of a complex system can be divided into distinct units, called modules.  The emergence of modules in biology appears to have resulted from the selection for efficient structures during evolution in a dynamic environment \cite{sun2007}.  Higher modularity is linked to optimized function and robustness to perturbation \cite{sun2007,variano2004}, however lower modularity is more advantageous over longer timescales, as it does not constrain the system to a rigid configuration \cite{park2015}.  The concept of modularity has been valuable to studying biological structure and function at various scales, including metabolic circuits \cite{ravasz2002}, antibody immune response to influenza \cite{bonomo2019}, protein-protein interaction networks \cite{mihalik2011}, ecological food webs \cite{krause2003}, and human brain networks \cite{sporns2016}.

In brain networks, nodes are typically defined by brain regions and edges can be based on either anatomical connections or relationships between the functional activity of different regions. Functional activity is a time series signal that can be acquired through various neuroimaging modalities, such as functional magnetic resonance imaging (fMRI), while subjects are at rest or performing specific tasks.  Based on the architecture of the resulting network, brain regions that are densely connected can be grouped into modules.  Modularity quantifies the overall community structure \citep{newman2006}.  The composition of modules has been used as a biomarker for illness \cite{chavez2010}, and among healthy subjects, individual differences in modularity correlated with individual differences in cognitive task performance \cite{yue2017,lebedev2018,chen2015}. Modularity has also been used to quantify changes in the brain during learning \cite{bassett2011} and to investigate organization of the functional network under specific demands, such as during visual tasks \cite{zhuo2011}. 

The use of modularity to distinguish how individual subjects' brain networks reconfigure during varying task demands has not been widely explored.  An important application is to non-pharmaceutical cognitive interventions, such as music therapy \cite{stuckey2010}, that are meant to enhance traditional medical treatments for patients of neurological disease and trauma. Significant network reconfiguration, quantifiable by the change in modularity, while subjects participate in these therapy enrichments may have implications for encouraged neuroplasticity and cognitive recovery. The theoretical grounding of modularity in biology may also shed light on why certain patients are more receptive than others to music-based interventions \cite{grimm2018} and why differences in the subtleties of the intervention, such as an auditory enrichment using music versus speech \cite{sarkamo2008}, significantly affect outcomes.

In this Rapid Communication, we investigate the modularity of whole-brain functional connectivity networks from fMRI data while healthy subjects listened to auditory pieces that varied in emotivity and cultural familiarity. We also introduce a ``super-module'' analysis method to study the consistency of module composition across different auditory pieces. The degree of modular structure in these networks during a subject's self-selected song is shown to be predictive of how the network architecture changes during familiar versus unfamiliar pieces. Namely, by classifying subjects into high and low modularity groups, we find that the low modularity networks exhibit significant adaptations during both familiar and unfamiliar music and speech; whereas the high modularity networks only significantly adapt during the unfamiliar pieces. We also find that coordinated activity among brain regions associated with self-referential thoughts is more consistent for the subjects in the high modularity group than it was for the low modularity group; whereas the module of auditory processing brain regions was more stable for subjects in the low modularity group. These results demonstrate the use of modularity as a viable quantifier of neural responses to music and speech. This work paves the way for understanding the diversity of responses patients of neurological disease or trauma may have to auditory-based therapy enrichments.

\section{Methods}
\label{sec:methods}

\subsection{fMRI Auditory Task}
\label{sec:pieces}
During fMRI, six auditory pieces from a pilot study \cite{karmonik2016,karmonik2020} were played for subjects (Table~\ref{tab:nSubj}).
We refer the reader to the Supplemental Material \cite{supp} for details about the cohort, fMRI acquisition, and fMRI pre-processing.

\paragraph{Self-Selected Song (Self)} 
Participants each chose a song to which they felt a strong emotional attachment.

\paragraph{Invention No.\ 1 (Bach)}
This piano piece in C major composed by J.\ S.\ Bach is representative of classical music that is culturally familiar to the participants in this study.  It compromises sufficient rhythmic and melodic variation to encourage engaged listening.

\paragraph{Jussuiraku (Gagaku)}
This instrumental gagaku piece from a Japanese opera in the oshiki-cho scale contains irregular rhythms, expressive noises, and deliberate detuning, and it is meant to contrast the piece by Bach. Gagaku is classical Japanese court music that was culturally unfamiliar to the participants in this study.

\paragraph{Xhosa Speech (Xhosa)}
Xhosa is a tonal Bantu language spoken in South Africa that contains three types of percussive click sounds.  The words and clicks are very distinct from sounds common to English and related languages, and therefore this speech excerpt is culturally unfamiliar to the participants.

\paragraph{Newscast Reading (Cronkite)}
This is a dry newscast presented by Walter Cronkite in 1973 about potential alien sightings.  Cronkite delivers the report dispassionately using a standard broadcasting speech pattern.

\paragraph{``The Great Dictator'' Speech (Chaplin)}
This is an emotionally-charged speech delivered by actor Charlie Chaplin while impersonating a dictator in his political satire film. The excerpt is meant to contrast Cronkite.

\begin{table}[h!]
\centering
\caption{Number of subjects that listened to each auditory piece overall and in the high and low modularity groups.}
\begin{tabular}{l | c | c c c c c c}
\hline
\hline
 & \textbf{Total} & Self & Bach & Gagaku & Xhosa & Cronkite & Chaplin  \\
\hline
All 	& \textbf{24} & 24 & 24 & 15 & 13 & 11 & 10  \\
\hline
Low $M_{\text{self}}$ & \textbf{9} & 9 & 9 & 8 & 6 & 6 & 7  \\
High $M_{\text{self}}$ & \textbf{15} & 15 & 15 & 7 & 7 & 5 & 3 \\
\hline
\hline
\end{tabular}
\label{tab:nSubj}
\end{table}

\subsection{Network construction}
\label{sec:network}

To construct functional activity networks, 84 Brodmann area (BA) brain regions are used as nodes, and the edges are determined by correlations in the activity between BAs during each auditory piece.  The Pearson correlation coefficient is computed between the time series of each BA pair to generate a weighted connectivity matrix for each subject listening to each auditory piece.  The functional connectivity matrix is binarized to a network density of 11.5\%, where the 400 edges with the highest weights are projected to unity and all others set to zero.  This density ensures the network is fully connected yet sufficiently sparse to improve the signal-to-noise ratio \cite{chen2015,yue2017}.
The resulting connectivity matrices are symmetric networks with unweighted, undirected edges.

\subsection{Modularity analysis}
\label{sec:modularity}

We use Newman's algorithm \cite{newman2006} as implemented in \cite{chen2015} to partition BAs into modules $\phi_k$, such that the arrangement maximizes modularity defined as
\begin{equation}
M(\{ \phi \}) = \frac{1}{2L}\sum_{k}\sum_{ij \in \phi_{k}}\big(A_{ij}-\frac{a_{i}a_{j}}{2L}\big),
\label{eq:M}
\end{equation}
where $L$ is the number of edges, $A_{ij}$ is the binarized connectivity matrix entry for BAs $i$ and $j$, and $a_i$ is the degree of BA $i$. The inner sum is evaluated for all $ij$ node pairs in module $\phi_k$.  The algorithm evaluates the modularity of each distribution of nodes into modules $\{ \phi \}$ against a null model, $a_{i}a_{j}/2L$, such that the existence of each intramodule link is scaled by the probability that a link between nodes $i$ and $j$ would be expected in a random network with the same degree distribution.  This is important because fluctuations in random networks have the potential to produce high modularity values \cite{guimera2004}.  

To quantify the adaptability of the functional network during different auditory processing demands, we consider the modular architecture during Self as a subject-specific baseline.  The amount the network architecture changes during other auditory pieces then reflects the extent that listening to these other pieces perturbs the brain from its baseline processing configuration.  
Change in modularity for each auditory piece $n$ is calculated as $\Delta M_n = M_n - M_\mathrm{Self}$, and the statistical significance of $\Delta M_n$ is determined by computing $p$-values from one-sample, two-tailed $t$-tests using the Statistics and Machine Learning Toolbox\texttrademark{} in MATLAB.
For 17 of the 24 subjects, $M_\mathrm{Self}$ is either the highest or lowest of the auditory pieces that each of those subjects listened to, motivating the use of Self as a baseline network for calculating $\Delta M_n$.
Furthermore, the substantial subject-to-subject variation warrants the use of $\Delta M_n$ rather than absolute $M$ values to compare the cohort results for different auditory pieces.
Namely, the average modularity over all pieces for each subject shows a substantial range, from $M = 0.36 \pm 0.02$ to $M = 0.57 \pm 0.01$ \cite{supp}.  
Sixteen of the 24 subjects have an average modularity that is significantly different at $p < 0.05$ than at least one other subject, and four subjects are significantly different at $p < 0.01$ than at least one other subject, based on two-sample, two-tailed $t$-tests.

\subsection{Module composition analysis}
\label{sec:module}

To study which brain regions are being commonly grouped together into modules, the functional connectivity matrices for all subjects are averaged together for each of the six different auditory pieces. We keep the top $\leq400$ edges that are statistically significant as determined by a one-sample, two-tailed $t$-test for each edge.    
The average connectivity matrices are then binarized, and modularity is calculated with Eq.~\ref{eq:M}.  This yields a set of modules $\{ \phi \}$ for each of the six average networks.

Newman's algorithm arbitrarily assigns a label to each module that it finds, and this label is not consistent across the different networks, even if the module composition appears qualitatively comparable. We therefore introduce a method to quantitatively compare modules across different networks. First, we determine the similarities in BA membership by calculating the Jaccard index \cite{real1999} between all pairs ($i \neq j$) of modules $i$ and $j$ across all six networks,
\begin{equation}
J(\phi_i,\phi_j) = \frac{|\mathbf{N}_{\phi_i} \cap \mathbf{N}_{\phi_j} |}{|\mathbf{N}_{\phi_i} \cup \mathbf{N}_{\phi_j} |},
\label{eq:jacc}
\end{equation}
where $\mathbf{N}_{\phi_i}$ is the set of BA nodes in module $\phi_i$.  A similarity measure of $J=1$ refers to two modules in different networks that have an identical node composition. $J=0$ means the two modules are either in the same network, or they are in different networks and do not have any nodes in common. Second, a set of \emph{super}-modules $\{ \Phi \}$ are determined using Eq.~\ref{eq:M}. Here, the network nodes are modules $\phi_i$ and the edges between each $ij$ node pair are the $J(\phi_i,\phi_j)$ similarity coefficients. In other words, the combined $\phi_i$ modules across the six networks are grouped into super-modules $\Phi_k$ based on overlap in the $\phi_i$ modules' sets of BA nodes, $\mathbf{N}_{\phi_i}$. The $\Phi_k$ groupings are then used to assign consistent labels to these modules that are analogs across the networks of different auditory pieces. The $\phi_i$ modules assigned to super-module $k$ in auditory piece $n$ collectively become $\Phi_{kn}$, and the $\mathbf{N}_{\phi_i}$ are then amalgamated, such that $\mathbf{N}_{\Phi_{kn}}$ is the total set of BAs in super-module $\Phi_{kn}$.

To quantify how stable the composition of each super-module $\Phi_k$ is across all auditory pieces, we calculate
\begin{equation}
P_{\Phi_k} = \frac{\sum\limits_{n \neq m}{J(\Phi_{kn},\Phi_{km})}}{[S(S-1)]/2},
\label{eq:Pmod}
\end{equation}
where $S=6$ is the number of auditory pieces, and $J(\Phi_{kn},\Phi_{km})$ is the Jaccard index between auditory pieces $n$ and $m$. $P_{\Phi_k} = 1$ means that super-module $\Phi_k$ has an identical set of BAs in all auditory pieces, whereas $P_{\Phi_k} = 0$ means that $\Phi_k$ is only present in one piece.

High and low modularity group networks are created by averaging the functional connectivity matrices for all applicable subjects in that group for each auditory piece. This results in 12 average networks, with super-modules $\Phi_k$ determined from the aggregate $\phi_i$ modules for all of these networks using the same method described above. 

\begin{figure}[h!]
\textbf{A}~~\includegraphics[width=0.45\textwidth]{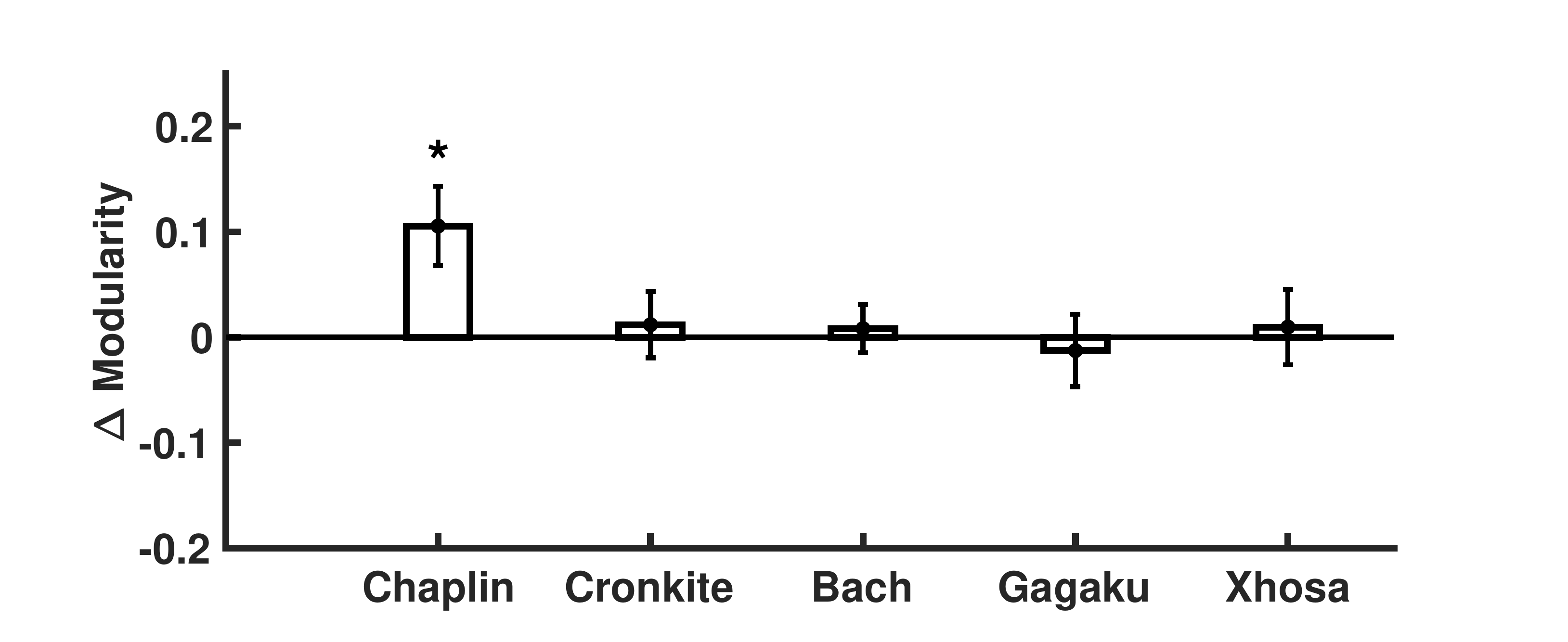} \\
\textbf{B}~\includegraphics[width=0.45\textwidth]{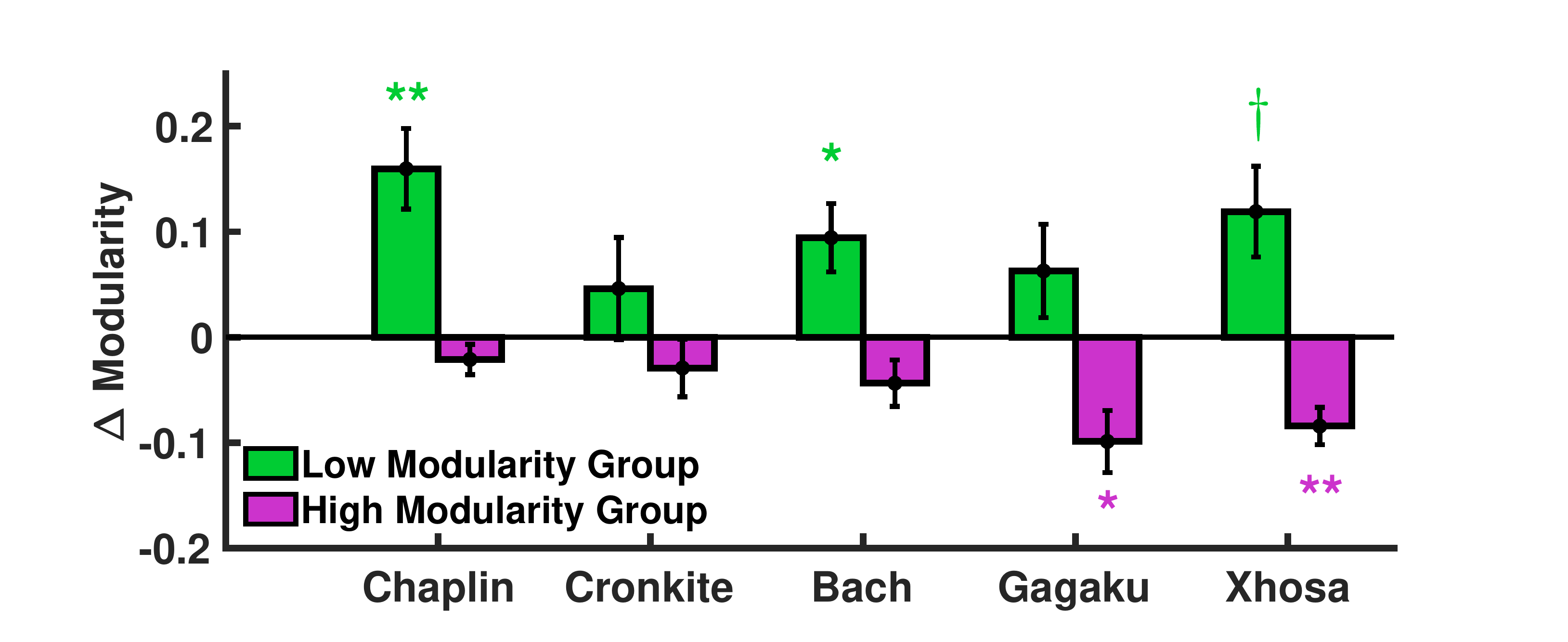}
\caption{Change in modularity across auditory pieces for \textbf{(A)} the cohort and \textbf{(B)} subjects divided into two groups. Error bars represent standard error.  Single asterisks indicate $p < 0.05$, double asterisks indicate $p < 0.01$, and the dagger indicates a marginal significance of $p = 0.053$.}
\label{fig:deltaM}
\end{figure}

\section{Results}
\label{sec:results}

\subsection{Network Adaptability}

The changes in modularity from Self to other auditory pieces are calculated for each subject individually and then averaged over all subjects for each auditory piece. There is a statistically significant increase in modularity during Chaplin (Fig.~\ref{fig:deltaM}A).  This suggests that there is some universality to having higher modularity for speech comprehension.  Indeed, a recent study that quantified functional network changes as the brain adapted to a speech listening task found that more successful listening was correlated with subjects having higher modularity during the task than during a resting state \cite{alavash2019}.  The increase in modularity that we observe also appears to be related to the emotive aspect of the Chaplin piece, since Cronkite did not elicit the same response.   

As mentioned above, modularity is either at its highest or lowest during Self for most subjects.  
We were interested in seeing if the null results for Cronkite, Bach, Gagaku, and Xhosa shown in Figure~\ref{fig:deltaM}A were due to the effects being cancelled out by these two different types of subjects.  To explore this, subjects are divided into low and high modularity groups based on if their modularity during their self-selected piece was lower or higher than the cohort average of $\overline{M}_\mathrm{Self} = 0.43$. Table~\ref{tab:nSubj} shows the number of subjects in each resulting group.  
The change in modularity is now averaged among subjects within each group (Fig.~\ref{fig:deltaM}B).  Subjects who have low modularity during Self adapt their network architecture during familiar (Chaplin and Bach) and unfamiliar pieces (Xhosa), whereas subjects who have high modularity during Self only significantly adapt during the unfamiliar pieces (Gagaku and Xhosa).  
Generalizing these results, they are in line with numerical experiments demonstrating that high modularity networks are more robust to perturbation \cite{variano2004}.  This has interesting implications for understanding why the effects of auditory-based therapeutic interventions often vary strongly across patients \cite{grimm2018}, warranting future research.  Patients with lower modularity during a favorite song may be more receptive to any type of music or auditory enrichment, whereas patients with higher modularity may require unique and unfamiliar auditory stimuli to sufficiently perturb their brain networks and encourage neuroplasticity.

\subsection{Module Composition}

\begin{figure}[h!]
\includegraphics[width=0.45\textwidth]{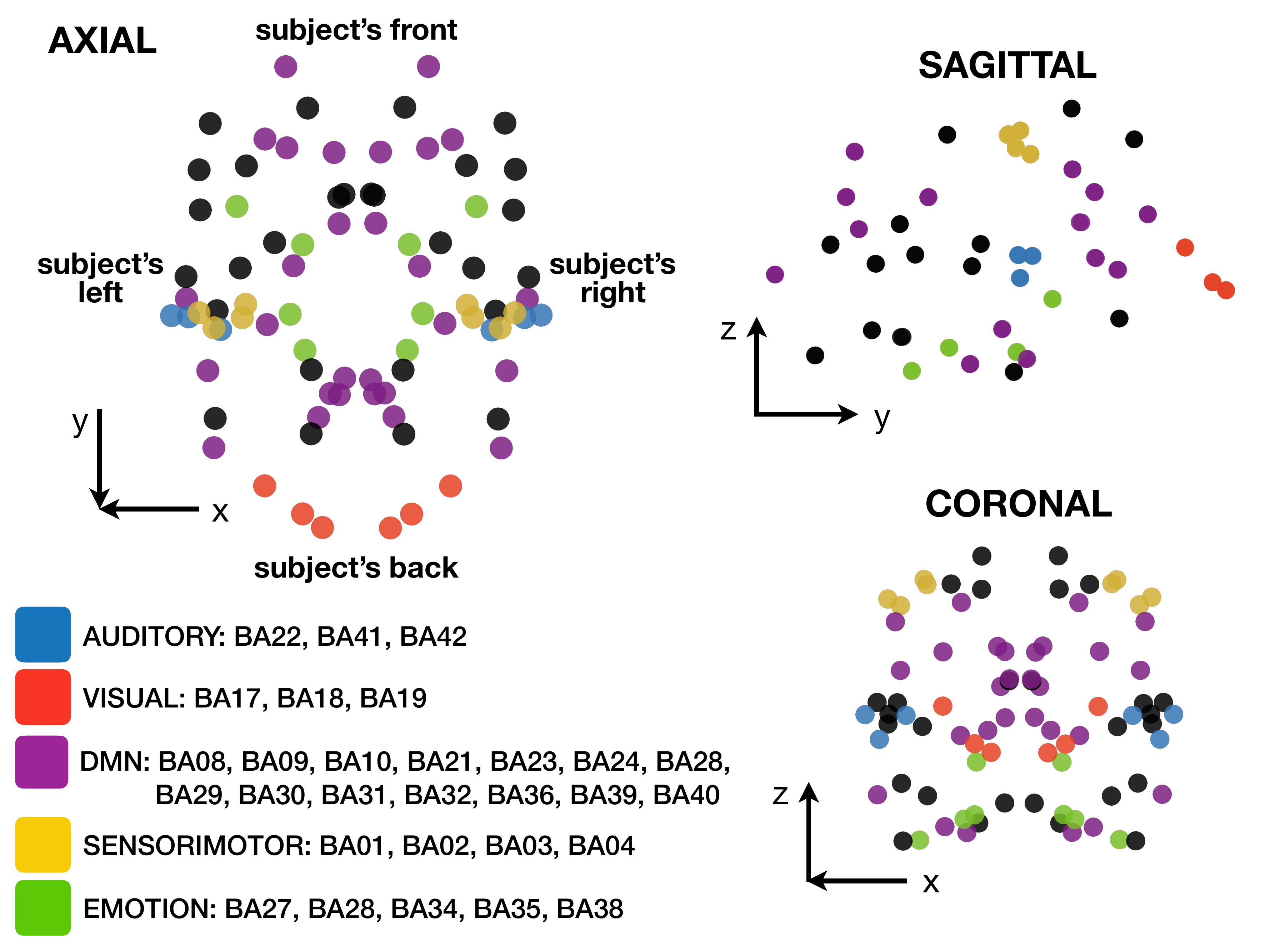}
\caption{The locations of BAs for relevant functions and the orientations of brain networks presented in Figures~\ref{fig:networkAVG} and \ref{fig:networkGroups}.}
\label{fig:key}
\end{figure}

We compare the module composition for the cohort and for the low and high modularity groups across all auditory pieces and look at the module memberships of the BAs associated with the following functions: auditory processing \cite{zilles2012}, visual and mental imagery processing \cite{zilles2012,ganis2004}, sensorimotor \cite{kaas2012}, emotion processing in the hippocampus \cite{geyer2013} and temporal pole \cite{olson2007}, and the default mode network (DMN) \cite{raichle2001,thatcher2014} (Fig.~\ref{fig:key}). Due to high subject-to-subject variation in edges when averaging brain networks across subjects, part of the hippocampus was often not assigned to a module. This low consistency of the hippocampus module allegiance across subjects is in agreement with prior brain modularity work \cite{wilkins2014}.

\begin{figure*}
\includegraphics[width=\textwidth]{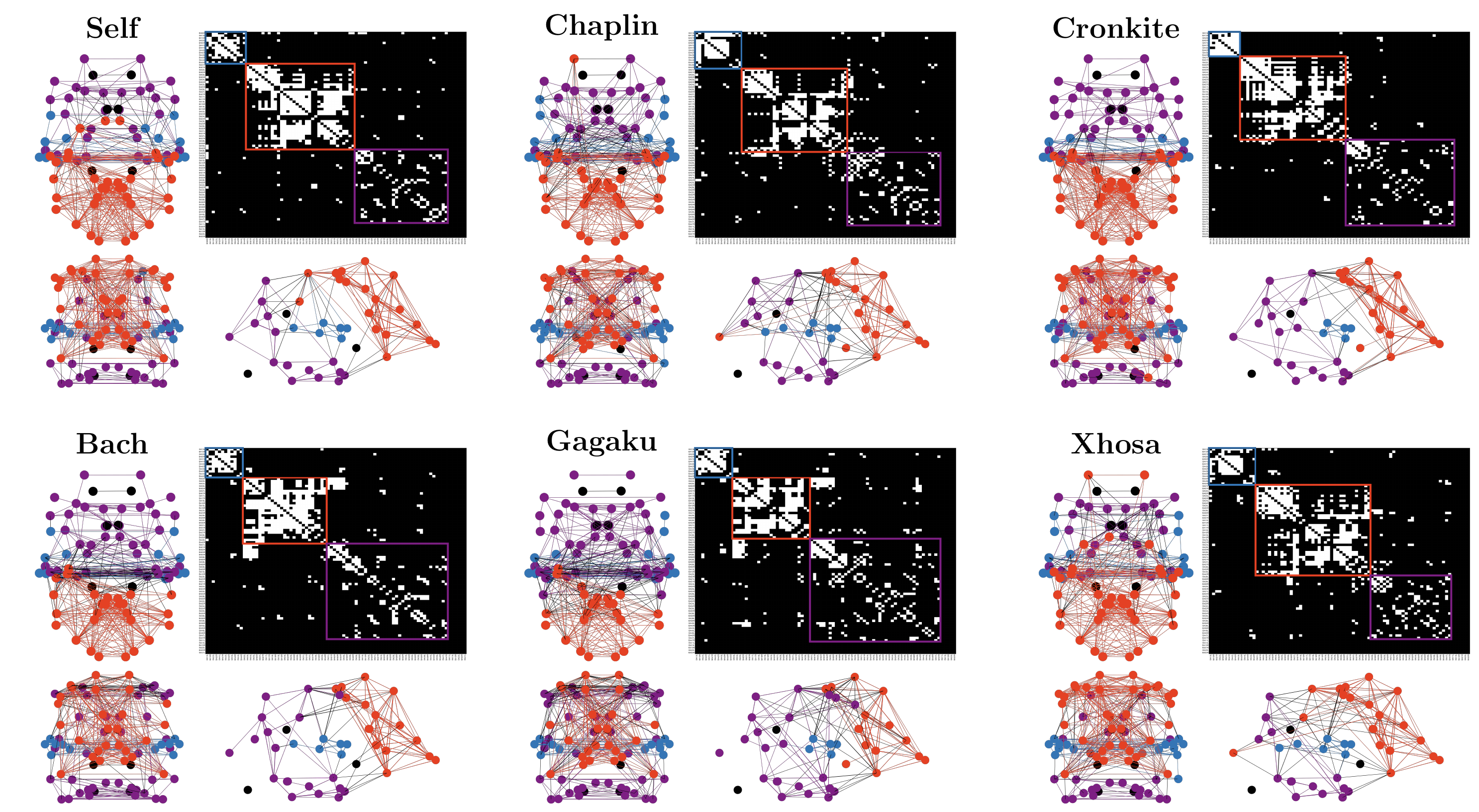}
\caption{Functional connectivity matrices averaged over all subjects during each auditory piece and brain networks oriented as in Fig.~\ref{fig:key}. The super-modules across all pieces are characterized by the auditory processing BAs ($\Phi_1$, blue), visual processing BAs ($\Phi_2$, red), and BAs in the DMN ($\Phi_3$, purple). Intermodule edges and BAs not assigned to a module are colored black.}
\label{fig:networkAVG}
\end{figure*}

Our analysis method described in Sec.~\ref{sec:module} identifies three super-modules (Fig.~\ref{fig:networkAVG}).  $\Phi_1$ contains the auditory processing BAs, $\Phi_2$ contains the visual processing BAs, and $\Phi_3$ contains BAs from the DMN. Individual BA module allegiance is listed in \cite{supp}. These super modules are fairly stable across all auditory pieces (Table~\ref{tab:Pmod}). The division of brain regions into functionally significant modules is in line with previous work that found that the task-based modular organization of brain regions is consistent with the regions needed to complete the task \cite{zhuo2011}.

\begin{table}[h!]
\centering
\caption{Stability of super-modules, $P_{\Phi_k}$, across each set of average networks.}
\begin{tabular}{l | c | c | c }
\hline
\hline
{\bf Super-module} & All Subjects & Low $M_{\text{self}}$ & High $M_{\text{self}}$ \\
\hline
{\bf $\Phi_1$ Auditory}	& $0.80 \pm 0.02$	& $0.76 \pm 0.03$	& $0.65 \pm 0.03$ \\
{\bf $\Phi_2$ Visual}	& $0.80 \pm 0.02$	& $0.84 \pm 0.02$	& $0.85 \pm 0.02$ \\
{\bf $\Phi_3$ DMN}	& $0.76 \pm 0.02$	& $0.31 \pm 0.03$	& $0.56 \pm 0.02$ \\
{\bf $\Phi_4$ Sensorimotor}	&  n/a	& $0.56 \pm 0.03$	& $0.60 \pm 0.05$ \\
{\bf $\Phi_5$ Emotion}	&  n/a	& $0.33 \pm 0.08$	& $0.42 \pm 0.09$ \\
\hline
\hline
\end{tabular}
\label{tab:Pmod}
\end{table}

\begin{figure*}[h!]
\includegraphics[width=\textwidth]{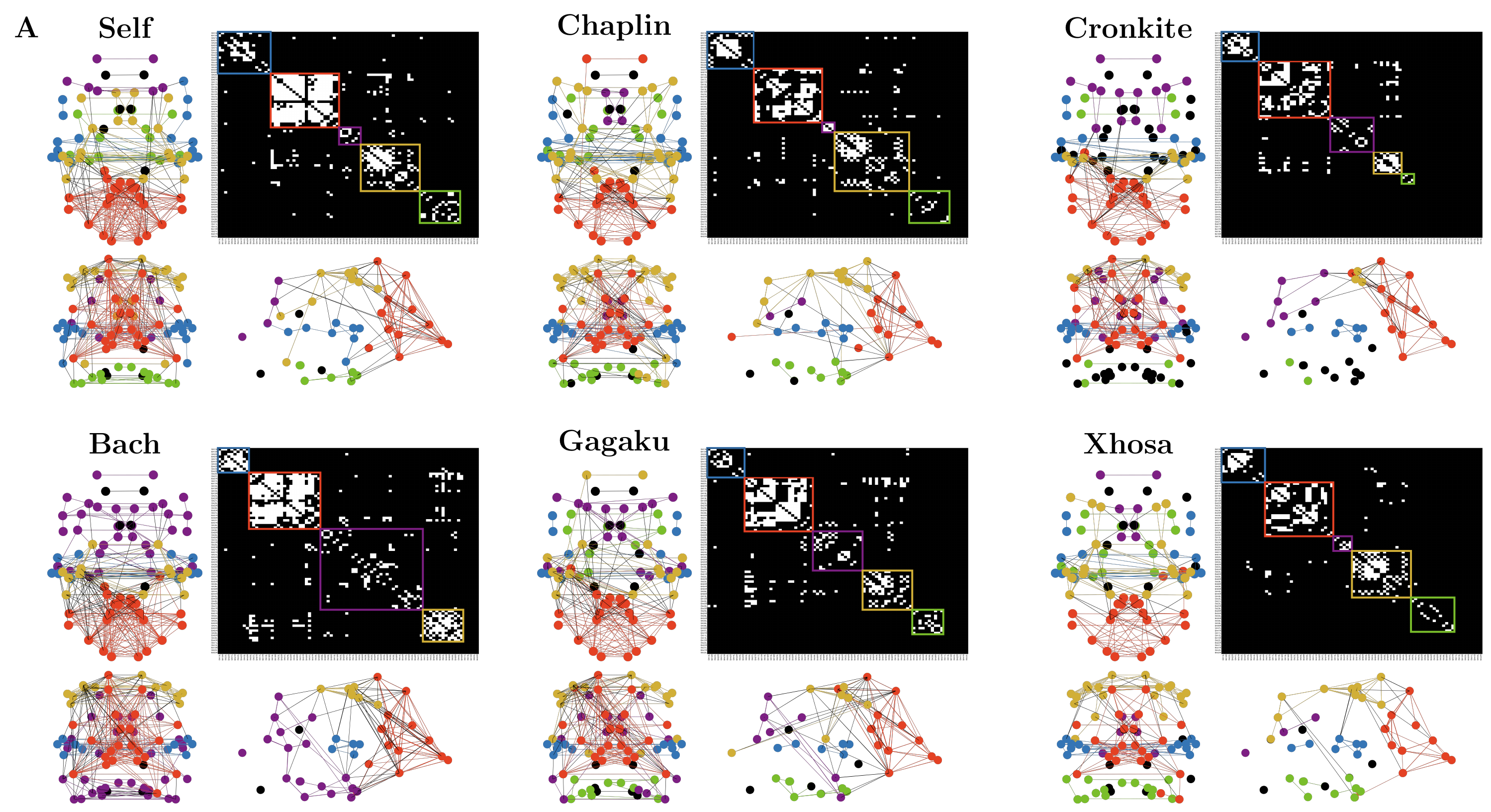}
\includegraphics[width=\textwidth]{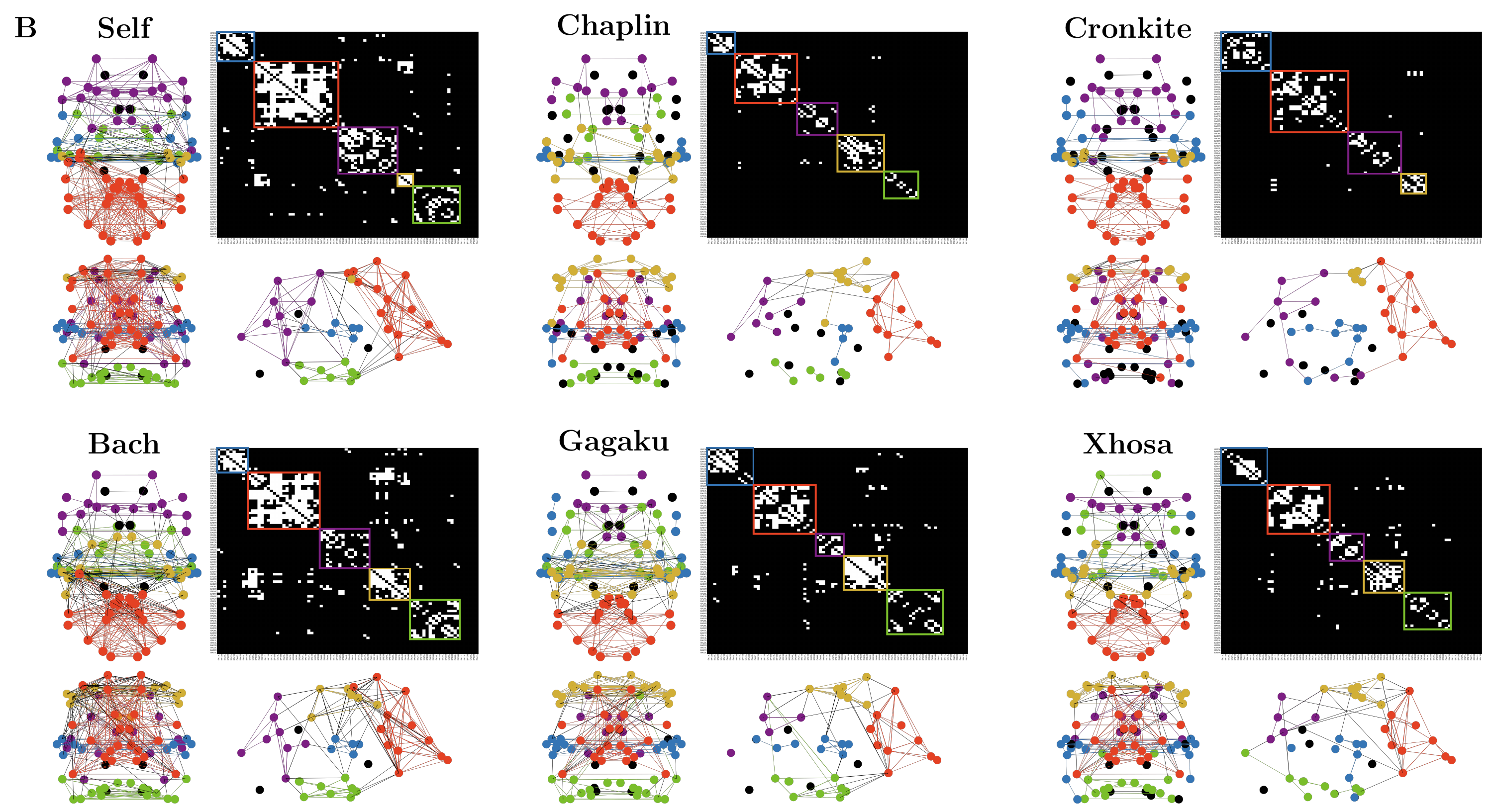}
\caption{Functional connectivity matrices averaged over all subjects in the {\bf (A)} low and {\bf (B)} high modularity groups during each auditory piece and brain networks oriented as in Fig.~\ref{fig:key}. The super-modules across all auditory pieces are characterized by BAs implicated in auditory processing ($\Phi_1$, blue), visual processing ($\Phi_2$, red), the DMN ($\Phi_3$, purple), sensorimotor function ($\Phi_4$, yellow), and emotion processing ($\Phi_5$, green). Intermodule edges and BAs not assigned to a module are colored black.}
\label{fig:networkGroups}
\end{figure*}

When dividing the subjects into low and high modularity groups, we identify two additional, functionally significant super-modules (Fig.~\ref{fig:networkGroups}): $\Phi_4$ contains the BAs involved in sensorimotor function and $\Phi_5$ was characterized by the emotion processing BAs. 
The more precise breakdown reveals group-wise differences in the stability of super-modules (Table~\ref{tab:Pmod}).
Namely, the DMN super-module ($\Phi_3$) was significantly more dynamic across the different auditory pieces for the low modularity group than the high modularity group. The DMN characterizes a set of brain regions that are active during stimulus-independent thought and have been linked to autobiographical memory and prospection \cite{raichle2001,spreng2010}. The fact that this super-module is more intact for the high modularity group could point to differences in how much subjects in the two groups engage in mind-wandering during the varying auditory task demands. In addition, the auditory super-module ($\Phi_1$) was moderately more stable across the different pieces for the low modularity group. It is interesting that while this group's community structure is overall more dynamic regardless of the familiarity of the stimulus (Fig.~\ref{fig:deltaM}B), on average there is this core auditory processing module. In the context of prior experiments and theory showing that lower modularity networks are better suited for performing complex tasks (i.e., requiring multiple types of cognitive functions) whereas higher modularity networks are more beneficial for fast response to straightforward tasks \cite{chen2015,yue2017,lebedev2018}, our results here may suggest that the low modularity group is optimizing both properties. That is, the low modularity group retains high fidelity of the $\Phi_1$ super-module for efficient processing of basic auditory features, but the overall network has high adaptability to process the additional cognitive components of the stimulus (e.g., familiarity, emotion, self-referential thoughts, memory).

\section{Conclusion}

In summary, we investigated the dynamic, whole-brain networks of subjects listening to music and speech through the lens of modularity. 
While many task-based neuroimaging studies focus on interpreting brain activations in the specific functional regions of interest (e.g., only those in the auditory cortex), whole-brain methods are poised to investigate how those activations fit into the larger context of the brain's comprehension of (auditory) information \cite{deWit2016}.
Furthermore, though a battery of graph theoretical measures are often used to quantify functional networks, modularity is a particularly elegant measure that has a biophysical grounding to study what drives a particular network reorganization \citep{lorenz2011}.
We have shown that baseline modularity and the familiarity of the stimulus both played a role in (1) the extent to which the brain network was perturbed, and (2) which groups of BAs across the whole-brain exhibited coordinated activity during the duration of the auditory pieces.
Even though we had a unitary, healthy population, our work highlighted the importance of considering results on a more individual level, as only considering the results for the cohort together averaged out the interesting group-wise differences. 
The trends seen for individuals with higher or lower modularity during their self-selected musical piece provided insight into the diversity of music and speech perception among people that might explain why the effect of a music intervention can vary strongly across individual patients. 
By demonstrating modularity as a quantifier of an individual's ``fingerprint'' \cite{finn2015} during general auditory processing and of the dynamic reorganization of the functional connectivity network during music and speech perception, this work may inform auditory-based interventions for patients of neurological disease and trauma.

\begin{acknowledgments}
The authors thank M.\ W.\ Deem for helpful discussions about the theory of this paper.  This work was supported by the Center for Theoretical Biological Physics at Rice University (National Science Foundation, PHY 1427654), the Ting Tsung and Wei Fong Chao Foundation, and the Houston Methodist Center for Performing Arts Medicine.
\end{acknowledgments}

\bibliography{Bonomo}
\end{document}


\singlespacing


\title{{\Large Supplemental Material}\\ 
\vspace{7mm}
Modularity allows classification of human brain networks during \\ music and speech perception}



\author{Melia E. Bonomo}
 
\author{Christof Karmonik}

\author{Anthony K. Brandt}

\author{J. Todd Frazier}




\begin{center}

\noindent{\Large Supplemental Material}\\ 
\noindent {\bf\large Modularity allows classification of human brain networks \\
during music and speech perception}

\vspace{3mm}
\noindent Melia E. Bonomo, Christof Karmonik, Anthony K. Brandt, J. Todd Frazier

\end{center}

\vspace{-25mm}

\noindent{\bf I. Additional Methods}

\noindent\emph{Participants.}

The study protocol was approved by the Houston Methodist Hospital Institutional Review Board, and all participants gave informed consent. Twenty-five healthy volunteers between the ages of 18 and 82 were recruited from the Houston community to participate in this study.  
Data for the first 12 subjects were previously collected during a pilot study \cite{karmonik2016,karmonik2020}.  
Participants were not taking any chronic medication or psychoactive drugs.  There was a heterogeneous distribution of gender, age, and extent of music education to avoid biasing to any of these factors.  Due to technical difficulty, data from one participant were excluded in the analysis.


\noindent\emph{MRI Acquisition.}

\begin{figure}[H]
\includegraphics[width=\linewidth]{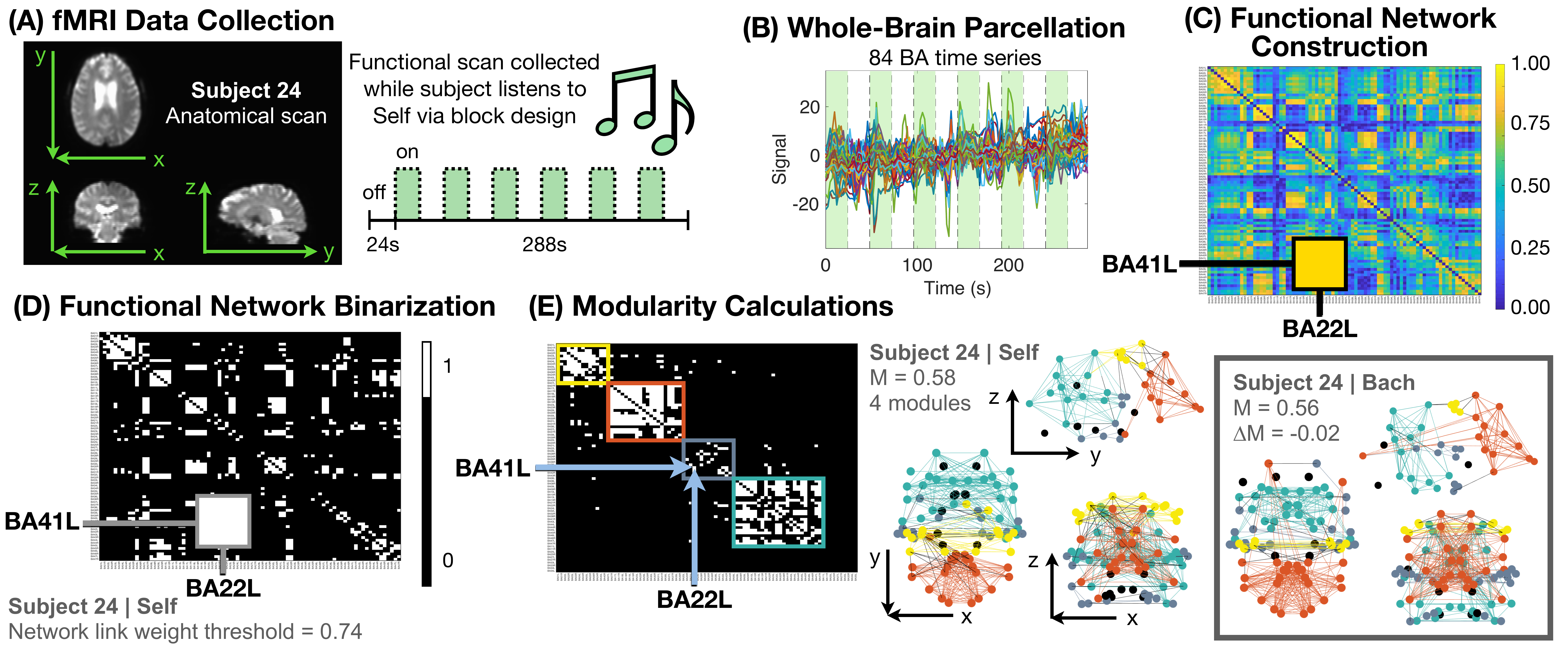}
\caption{Protocol followed for processing MRI data. Analysis of Subject 12 listening to Bach is shown as an example. 
{\bf (A)} The anatomical MRI scan is collected and shown here using the radiological convention, where $-x$ is right, $+x$ is left, $-y$ is anterior, $+y$ is posterior, $-z$ is inferior, and $+z$ is superior. The functional scan is collected during a 312s run that follows a block design for each auditory piece.  The anatomical and functional scans are aligned to obtain the BOLD signal from each 3mm voxel over the whole brain.
{\bf (B)} The whole brain is parcellated into 84 BA regions, and the BOLD signal is averaged over all voxels within each region. The functional activity of each BA over time is shown here, where the green shaded bars indicate when the auditory stimulus was on.
{\bf (C)} A functional connectivity matrix of 84x84 BAs is generated by calculating pairwise correlations between all time series. The matrix axes are ordered BA01L, BA01R, BA02L, BA02R, etc.\ for left (L) and right (R) hemisphere BAs. The correlation between the signal in BA41L and BA22L, both involved in auditory processing, is highlighted as an example. A complete list of BAs used in this analysis is provided in Tables S1, S2, and S3.
{\bf (D)} The top 11.5\% of edges of the functional connectivity matrix are set to 1 and all other edges are set to 0.  In this example, keeping the top 11.5\% of edges meant setting a correlation coefficient threshold of 0.74. The binarized matrix axes are ordered as in {\bf C}.
{\bf (E)} Modularity is calculated using Newman's algorithm. 
The functional connectivity matrix entries are rearranged here to visualize the BA composition of each of four modules. To visualize the network, BA network node coordinates are extracted from AFNI, and edges are constructed from the binarized connectivity matrix.  Intra-module connections and nodes are color-coded by module, and inter-module connections are black.}
\label{fig:protocol}
\end{figure}

Neuroimaging took place at the Houston Methodist Research Institute MRI core using a Philips Ingenia 3.0T scanner.  Anatomical scans were acquired with a turbo field echo pulse sequence at an 8.2ms repetition time and 3.8ms echo time (field of view of 24~x~24~x~16.5cm, 1.0mm isotropic resolution, axial orientation).  Functional scans were acquired in T2* weighted slices with an echo planar imaging pulse sequence at a 2400ms repetition time and 35ms echo time (field of view: 22~x~22~x~12cm, resolution: 1.5~x~1.5~x~3.0mm, axial orientation).
The functional imaging was obtained while subjects listened to each auditory piece through headphones in the scanner bed. High frequencies were increased during playback of each audio track using the iTunes digital equalizer to account for attenuation of these tones in the air tubing used to connect to the headphones.  The listening task followed a standard block design, in which there was silence for 10 brain volumes (24s), followed by 12 blocks alternating 10-volume intervals of auditory stimulus and silence, for a total of 130 volumes (312s) in each run (see Figure~\ref{fig:protocol}). The order of pieces played was Self, Bach, Gagaku, Xhosa, Cronkite, and Chaplin. The Self songs were downloaded from iTunes (Apple Inc). The number of pieces that each subject listened to was dependent on how long they were comfortable staying in the scanner.


\noindent\emph{MRI Pre-Processing.}

The MRI data underwent standard pre-processing in AFNI \cite{cox2012} for alignment of the anatomical and functional scans, motion correction, spatial smoothing, and bandpass filtering of the blood oxygen level-dependent (BOLD) signal to remove constant offset and high-frequencies.  The AFNI software was also used to transform the data into Talairach space and reconstruct the whole-brain signal into 84 Brodmann areas (BAs), in which the time series were averaged over all voxels segmented into each BA.  Previous work has shown consistency in modularity trends across different parcellation atlases \cite{yue2017}.  The first 24s of silence during each run were not included in the analysis.






\newpage
\noindent{\bf II. Modularity Results for Individual Subjects}

\begin{figure}[H]
{\bf A}\includegraphics[width=0.9\linewidth]{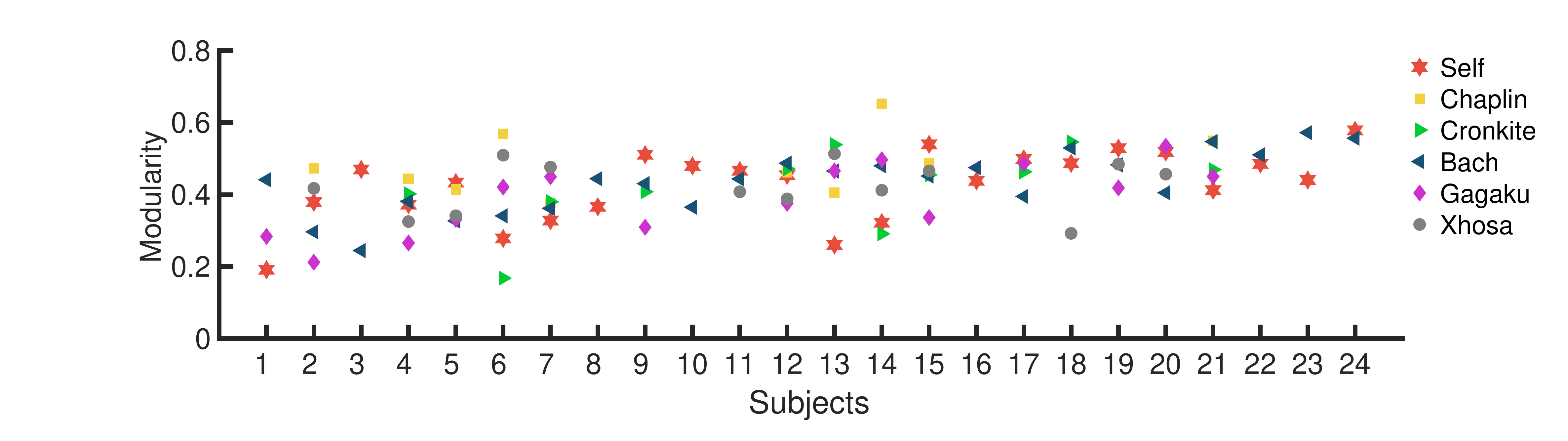} \\
{\bf B}\includegraphics[width=0.9\linewidth]{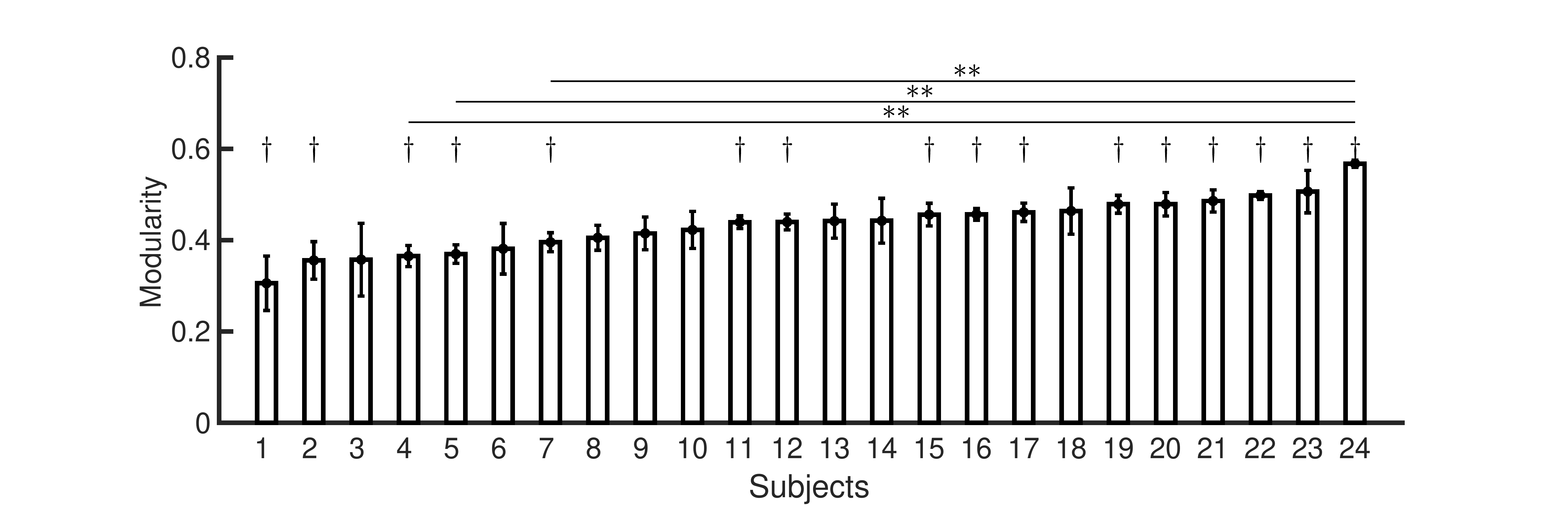}
\caption{Differences in modularity among individual subjects. \textbf{(A)} Modularity for each subject as they listened to each auditory piece.  Subjects are ordered from low to high average modularity. Nine of the 24 subjects exhibit decreased network modularity during all of the other auditory pieces from what it was during Self, and eight exhibit increased modularity during all of the other auditory pieces. \textbf{(B)} Average modularity over all auditory pieces for each subject. Subjects are ordered as in {\bf A}. Error bars are standard error. The daggers indicate $p < 0.05$ between that subject and at least one other subject.  The double asterisks indicate $p < 0.01$ between the specified subjects. Statistics are computed based on two-sample, two-tailed $t$-tests.}
\label{fig:Individuals}
\end{figure}




\newpage
\noindent{\bf III. Module Membership of Brodmann Areas}

\vspace{2mm}
\noindent Tables S1, S2, S3 show the super-module assignments for each BA during each auditory piece for the average networks created with all subjects, low modularity subjects, and high modularity subjects, respectively. BAs are ordered based on their structure number. 
This method is generally able to place small modules that are highly isolated in one network into an appropriate super-module for cross-network analyses; however, NaN means that there were no links connected to that BA when the average network was created and/or a super-module could not be assigned. The BAs most often not assigned to a module due to high subject-to-subject variation in edges are the orbital frontal cortex (BA 11), the anterior cingulate (BA 33), and part of the hippocampus (BA 27)

\vspace{10mm}
\begin{longtable}[H]{ll|cccccc}
\caption{Modules for Average Networks of All Subjects} \\
\hline
\textbf{ } & \textbf{Brodmann Area Name}         & \textbf{Self} & \textbf{Chaplin} & \textbf{Gagaku} & \textbf{Bach} & \textbf{Cronkite} & \textbf{Xhosa} \\
\hline
\endfirsthead
\hline
\textbf{ } & \textbf{Brodmann Area Name}         & \textbf{Self} & \textbf{Chaplin} & \textbf{Gagaku} & \textbf{Bach} & \textbf{Cronkite} & \textbf{Xhosa} \\
\hline
\endhead
BA01L       & Primary somatosensory   cortex             & 2             & 2             & 3               & 3              & 2                & 2                 \\
BA01R       & Primary somatosensory   cortex             & 2             & 2             & 3               & 3              & 2                & 2                 \\
BA02L       & Secondary somatosensory cortex             & 2             & 2             & 3               & 2              & 2                & 2                 \\
BA02R       & Secondary   somatosensory cortex           & 2             & 2             & 3               & 3              & 2                & 2                 \\
BA03L       & Tertiary   somatosensory cortex            & 2             & 2             & 2               & 2              & 2                & 2                 \\
BA03R       & Tertiary   somatosensory cortex            & 2             & 2             & 3               & 3              & 2                & 2                 \\
BA04L       & Primary motor cortex                       & 2             & 2             & 2               & 2              & 2                & 2                 \\
BA04R       & Primary motor cortex                       & 2             & 2             & 3               & 3              & 2                & 2                 \\
BA05L       & Superior parietal   sulcus                 & 2             & 2             & 2               & 2              & 2                & 2                 \\
BA05R       & Superior parietal   sulcus                 & 2             & 2             & 2               & 2              & 2                & 2                 \\
BA06L       & Supplementary motor   area                 & 2             & 3             & 3               & 3              & 3                & 2                 \\
BA06R       & Supplementary motor   area                 & 1             & 3             & 3               & 3              & 3                & 2                 \\
BA07L       & Superior parietal   gyrus                  & 2             & 2             & 2               & 2              & 2                & 2                 \\
BA07R       & Superior parietal   gyrus                  & 2             & 2             & 2               & 2              & 2                & 2                 \\
BA08L       & Pre-supplementary   motor area             & 3             & 3             & 3               & 3              & 3                & 3                 \\
BA08R       & Pre-supplementary   motor area             & 3             & 3             & 3               & 3              & 3                & 3                 \\
BA09L       & Dorsolateral   prefrontal cortex           & 3             & 3             & 3               & 3              & 3                & 3                 \\
BA09R       & Dorsolateral   prefrontal cortex           & 3             & 3             & 3               & 3              & 3                & 3                 \\
BA10L       & Fronto-parietal   cortex                   & 3             & 2             & 3               & 3              & 3                & 2                 \\
BA10R       & Fronto-parietal   cortex                   & 3             & 3             & 3               & 3              & 3                & 2                 \\
BA11L       & Orbital frontal   cortex                   & NaN           & NaN           & NaN             & NaN            & NaN              & NaN               \\
BA11R       & Orbital frontal   cortex                   & NaN           & NaN           & NaN             & NaN            & NaN              & NaN               \\
BA13L       & Insula                & 1             & 1             & 1               & 1              & 1                & 1                 \\
BA13R       & Insula                & 1             & 1             & 1               & 1              & 1                & 1                 \\
BA17L       & Primary visual cortex                      & 2             & 2             & 2               & 2              & 2                & 2                 \\
BA17R       & Primary visual cortex                      & 2             & 2             & 2               & 2              & 2                & 2                 \\
BA18L       & Secondary visual   cortex                  & 2             & 2             & 2               & 2              & 2                & 2                 \\
BA18R       & Secondary visual   cortex                  & 2             & 2             & 2               & 2              & 2                & 2                 \\
BA19L       & Cuneus                & 2             & 2             & 2               & 2              & 2                & 2                 \\
BA19R       & Cuneus                & 2             & 2             & 2               & 2              & 2                & 2                 \\
BA20L       & Inferior temporal   gyrus                  & 3             & 3             & 3               & 3              & 3                & 3                 \\
BA20R       & Inferior temporal   gyrus                  & 3             & 3             & 3               & 3              & 3                & 3                 \\
BA21L       & Medial temporal gyrus                      & 3             & 3             & 3               & 3              & 3                & 3                 \\
BA21R       & Medial temporal gyrus                      & 3             & 1             & 3               & 3              & 3                & 1                 \\
BA22L       & Superior temporal   gyrus                  & 1             & 1             & 1               & 1              & 1                & 1                 \\
BA22R       & Superior temporal   gyrus                  & 1             & 1             & 1               & 1              & 1                & 1                 \\
BA23L       & Posterior cingulate   cortex1              & 2             & 2             & 2               & 2              & 2                & 2                 \\
BA23R       & Posterior cingulate   cortex1              & 2             & 2             & 2               & 2              & 2                & 2                 \\
BA24L       & Dorsal anterior   cingulate cortex         & 2             & 3             & 3               & 3              & 3                & 2                 \\
BA24R       & Dorsal anterior   cingulate cortex         & 2             & 3             & 3               & 3              & 3                & 3                 \\
BA25L       & Subgenual anterior   cingulate cortex      & 3             & 3             & 3               & 3              & 3                & 3                 \\
BA25R       & Subgenual anterior   cingulate cortex      & 3             & 3             & 3               & 3              & 3                & 3                 \\
BA27L       & Parahippocampal   gyrus1                   & NaN           & 2             & 2               & NaN            & 2                & NaN               \\
BA27R       & Parahippocampal   gyrus1                   & NaN           & NaN           & NaN             & NaN            & NaN              & NaN               \\
BA28L       & Hippocampal area1                          & 3             & 3             & 3               & 3              & 3                & 3                 \\
BA28R       & Hippocampal area1                          & 3             & 3             & 3               & 3              & 3                & 3                 \\
BA29L       & Retrosplenial cortex1                      & 2             & 2             & 2               & 2              & 2                & 2                 \\
BA29R       & Retrosplenial cortex1                      & 2             & 2             & 2               & 2              & 2                & 2                 \\
BA30L       & Retrosplenial cortex2                      & 2             & 2             & 2               & 2              & 2                & 2                 \\
BA30R       & Retrosplenial cortex2                      & 2             & 2             & 2               & 2              & 2                & 2                 \\
BA31L       & Posterior cingulate   cortex2              & 2             & 2             & 2               & 2              & 2                & 2                 \\
BA31R       & Posterior cingulate   cortex2              & 2             & 2             & 2               & 2              & 2                & 2                 \\
BA32L       & Pregenual anterior   cingulate cortex      & 3             & 3             & 3               & 3              & 3                & 3                 \\
BA32R       & Pregenual anterior   cingulate cortex      & 3             & 3             & 3               & 3              & 3                & 3                 \\
BA33L       & Rostral anterior   cingulate cortex        & NaN           & NaN           & NaN             & NaN            & NaN              & NaN               \\
BA33R       & Rostral anterior   cingulate cortex        & NaN           & NaN           & NaN             & NaN            & NaN              & NaN               \\
BA34L       & Hippocampus           & 3             & 3             & 3               & 3              & 3                & 3                 \\
BA34R       & Hippocampus           & 3             & 3             & 3               & 3              & 3                & 3                 \\
BA35L       & Hippocampal area2                          & 3             & 3             & 3               & 3              & 3                & 3                 \\
BA35R       & Hippocampal area2                          & 3             & 3             & 3               & 3              & 3                & 3                 \\
BA36L       & Parahippocampal   gyrus2                   & 3             & 3             & 3               & 3              & 3                & 3                 \\
BA36R       & Parahippocampal   gyrus2                   & 3             & 3             & 3               & 3              & 2                & 3                 \\
BA37L       & Occipital-temporal   cortex                & 2             & 2             & 2               & 2              & 2                & 2                 \\
BA37R       & Occipital-temporal   cortex                & 2             & 2             & 2               & 2              & 2                & 2                 \\
BA38L       & Temporal pole         & 3             & 3             & 3               & 3              & 3                & 3                 \\
BA38R       & Temporal pole         & 3             & 3             & 3               & 3              & 3                & 3                 \\
BA39L       & Angular gyrus         & 2             & 2             & 2               & 2              & 2                & 2                 \\
BA39R       & Angular gyrus         & 2             & 2             & 2               & 2              & 2                & 2                 \\
BA40L       & Intra-parietal sulcus                      & 2             & 2             & 3               & 2              & 2                & 2                 \\
BA40R       & Intra-parietal sulcus                      & 2             & 2             & 3               & 2              & 2                & 2                 \\
BA41L       & Primary auditory   cortex                  & 1             & 1             & 1               & 1              & 1                & 1                 \\
BA41R       & Primary auditory   cortex                  & 1             & 1             & 1               & 1              & 1                & 1                 \\
BA42L       & Secondary auditory   cortex                & 1             & 1             & 1               & 1              & 1                & 1                 \\
BA42R       & Secondary auditory   cortex                & 1             & 1             & 1               & 1              & 1                & 1                 \\
BA43L       & Postcentral gyrus                          & 1             & 1             & 1               & 1              & 1                & 1                 \\
BA43R       & Postcentral gyrus                          & 1             & 1             & 1               & 1              & 1                & 1                 \\
BA44L       & Opercular part of   inferior frontal gyrus & 1             & 1             & 3               & 1              & 3                & 1                 \\
BA44R       & Opercular part of   inferior frontal gyrus & 1             & 1             & 1               & 1              & 3                & 1                 \\
BA45L       & Inferior frontal   gyrus                   & 3             & 1             & 3               & 3              & 3                & 1                 \\
BA45R       & Inferior frontal   gyrus                   & 3             & 1             & 1               & 3              & 3                & 1                 \\
BA46L       & Medial prefrontal   cortex                 & 3             & 3             & 3               & 3              & 3                & 3                 \\
BA46R       & Medial prefrontal   cortex                 & 3             & 3             & 3               & 3              & 3                & 3                 \\
BA47L       & Ventro-lateral   prefrontal cortex         & 3             & 3             & 3               & 3              & 3                & 3                 \\
BA47R       & Ventro-lateral   prefrontal cortex         & 3             & 3             & 3               & 3              & 3                & 3                \\
\hline
\end{longtable}

\vspace{10mm}

\begin{longtable}[H]{ll|cccccc} 
\caption{Modules for Low Modularity Group} \\
\hline
\textbf{ } & \textbf{Brodmann Area Name}         & \textbf{Self} & \textbf{Chaplin} & \textbf{Gagaku} & \textbf{Bach} & \textbf{Cronkite} & \textbf{Xhosa} \\
\hline
\endfirsthead
\hline
\textbf{ } & \textbf{Brodmann Area Name}         & \textbf{Self} & \textbf{Chaplin} & \textbf{Gagaku} & \textbf{Bach} & \textbf{Cronkite} & \textbf{Xhosa} \\
\hline
\endhead
BA01L       & Primary somatosensory   cortex             & 4             & 4             & 4               & 4              & 4                & 4                 \\
BA01R       & Primary somatosensory   cortex             & 4             & 4             & 4               & 4              & 4                & 4                 \\
BA02L       & Secondary   somatosensory cortex           & 4             & 4             & 4               & 4              & 4                & 4                 \\
BA02R       & Secondary   somatosensory cortex           & 4             & 4             & 4               & 4              & 4                & 4                 \\
BA03L       & Tertiary   somatosensory cortex            & 4             & 4             & 4               & 4              & 4                & 4                 \\
BA03R       & Tertiary   somatosensory cortex            & 4             & 4             & 4               & 4              & 4                & 4                 \\
BA04L       & Primary motor cortex                       & 4             & 4             & 2               & 4              & 2                & 4                 \\
BA04R       & Primary motor cortex                       & 4             & 4             & 4               & 4              & 4                & 4                 \\
BA05L       & Superior parietal   sulcus                 & 2             & 4             & 2               & 2              & 2                & 4                 \\
BA05R       & Superior parietal   sulcus                 & 4             & 4             & 4               & 4              & 4                & 4                 \\
BA06L       & Supplementary motor   area                 & 4             & 4             & 4               & 4              & 3                & 4                 \\
BA06R       & Supplementary motor   area                 & 4             & 4             & 4               & 4              & 3                & 4                 \\
BA07L       & Superior parietal   gyrus                  & 2             & 2             & 2               & 2              & 2                & 2                 \\
BA07R       & Superior parietal   gyrus                  & 2             & 2             & 2               & 2              & 2                & 2                 \\
BA08L       & Pre-supplementary   motor area             & 3             & 4             & 3               & 3              & 3                & 4                 \\
BA08R       & Pre-supplementary   motor area             & 3             & 4             & 3               & 3              & 3                & 4                 \\
BA09L       & Dorsolateral   prefrontal cortex           & 3             & 4             & 3               & 3              & 3                & 4                 \\
BA09R       & Dorsolateral   prefrontal cortex           & 3             & 4             & 3               & 3              & 3                & 4                 \\
BA10L       & Fronto-parietal   cortex                   & 3             & 2             & 4               & 3              & 3                & 3                 \\
BA10R       & Fronto-parietal   cortex                   & 3             & 2             & 4               & 3              & 3                & 3                 \\
BA11L       & Orbital frontal   cortex                   & NaN           & NaN           & NaN             & NaN            & NaN              & NaN               \\
BA11R       & Orbital frontal   cortex                   & NaN           & NaN           & NaN             & NaN            & NaN              & NaN               \\
BA13L       & Insula                & 1             & 1             & 1               & 1              & 1                & 1                 \\
BA13R       & Insula                & 1             & 1             & 1               & 1              & 1                & 1                 \\
BA17L       & Primary visual cortex                      & 2             & 2             & 2               & 2              & 2                & 2                 \\
BA17R       & Primary visual cortex                      & 2             & 2             & 2               & 2              & 2                & 2                 \\
BA18L       & Secondary visual   cortex                  & 2             & 2             & 2               & 2              & 2                & 2                 \\
BA18R       & Secondary visual   cortex                  & 2             & 2             & 2               & 2              & 2                & 2                 \\
BA19L       & Cuneus                & 2             & 2             & 2               & 2              & 2                & 2                 \\
BA19R       & Cuneus                & 2             & 2             & 2               & 2              & 2                & 2                 \\
BA20L       & Inferior temporal   gyrus                  & 5             & 5             & 3               & 3              & NaN              & 5                 \\
BA20R       & Inferior temporal   gyrus                  & 5             & 4             & 3               & 3              & NaN              & 2                 \\
BA21L       & Medial temporal gyrus                      & 1             & 5             & 3               & 3              & NaN              & 5                 \\
BA21R       & Medial temporal gyrus                      & 1             & 1             & 3               & 3              & NaN              & NaN               \\
BA22L       & Superior temporal   gyrus                  & 1             & 1             & 1               & 1              & 1                & 1                 \\
BA22R       & Superior temporal   gyrus                  & 1             & 1             & 1               & 1              & 1                & 1                 \\
BA23L       & Posterior cingulate   cortex1              & 2             & 2             & 2               & 2              & 2                & 2                 \\
BA23R       & Posterior cingulate   cortex1              & 2             & 2             & 2               & 2              & 2                & 2                 \\
BA24L       & Dorsal anterior   cingulate cortex         & 4             & 3             & 3               & 3              & 3                & 3                 \\
BA24R       & Dorsal anterior   cingulate cortex         & 4             & 3             & 3               & 3              & 3                & 3                 \\
BA25L       & Subgenual anterior   cingulate cortex      & 5             & 5             & 5               & 3              & NaN              & 5                 \\
BA25R       & Subgenual anterior   cingulate cortex      & 5             & 5             & 5               & 3              & NaN              & 5                 \\
BA27L       & Parahippocampal   gyrus1                   & 2             & 2             & NaN             & 2              & NaN              & NaN               \\
BA27R       & Parahippocampal   gyrus1                   & NaN           & NaN           & NaN             & NaN            & NaN              & NaN               \\
BA28L       & Hippocampal area1                          & 5             & 5             & 5               & 3              & NaN              & 5                 \\
BA28R       & Hippocampal area1                          & 5             & 5             & NaN             & 3              & NaN              & 5                 \\
BA29L       & Retrosplenial cortex1                      & 2             & 2             & 2               & 2              & 2                & 2                 \\
BA29R       & Retrosplenial cortex1                      & 2             & 2             & 2               & 2              & 2                & 2                 \\
BA30L       & Retrosplenial cortex2                      & 2             & 2             & 2               & 2              & 2                & 2                 \\
BA30R       & Retrosplenial cortex2                      & 2             & 2             & 2               & 2              & 2                & 2                 \\
BA31L       & Posterior cingulate   cortex2              & 2             & 2             & 2               & 2              & 2                & 2                 \\
BA31R       & Posterior cingulate   cortex2              & 2             & 2             & 2               & 2              & 2                & 2                 \\
BA32L       & Pregenual anterior   cingulate cortex      & 4             & 3             & 3               & 3              & 3                & 3                 \\
BA32R       & Pregenual anterior   cingulate cortex      & 4             & 3             & 3               & 3              & 3                & 3                 \\
BA33L       & Rostral anterior   cingulate cortex        & NaN           & NaN           & NaN             & NaN            & NaN              & NaN               \\
BA33R       & Rostral anterior   cingulate cortex        & NaN           & NaN           & NaN             & NaN            & NaN              & NaN               \\
BA34L       & Hippocampus           & NaN           & 5             & NaN             & 3              & NaN              & NaN               \\
BA34R       & Hippocampus           & 5             & 5             & 5               & 3              & NaN              & 5                 \\
BA35L       & Hippocampal area2                          & 5             & 5             & 5               & 3              & NaN              & 5                 \\
BA35R       & Hippocampal area2                          & 5             & 4             & 3               & 3              & NaN              & 5                 \\
BA36L       & Parahippocampal   gyrus2                   & 5             & 5             & 5               & 3              & NaN              & 5                 \\
BA36R       & Parahippocampal   gyrus2                   & 5             & 4             & 3               & 2              & NaN              & 2                 \\
BA37L       & Occipital-temporal   cortex                & 2             & 2             & 2               & 2              & 2                & 2                 \\
BA37R       & Occipital-temporal   cortex                & 2             & 4             & 2               & 2              & 2                & 2                 \\
BA38L       & Temporal pole         & 5             & NaN           & 5               & 3              & 5                & 5                 \\
BA38R       & Temporal pole         & 5             & 5             & 5               & 3              & 5                & 5                 \\
BA39L       & Angular gyrus         & 2             & 2             & 2               & 2              & 2                & 2                 \\
BA39R       & Angular gyrus         & 2             & 2             & 2               & 2              & 2                & 2                 \\
BA40L       & Intra-parietal sulcus                      & 4             & 4             & 4               & 4              & 2                & 4                 \\
BA40R       & Intra-parietal sulcus                      & 4             & 4             & 4               & 4              & 4                & 4                 \\
BA41L       & Primary auditory   cortex                  & 1             & 1             & 1               & 1              & 1                & 1                 \\
BA41R       & Primary auditory   cortex                  & 1             & 1             & 1               & 1              & 1                & 1                 \\
BA42L       & Secondary auditory   cortex                & 1             & 1             & 1               & 1              & 1                & 1                 \\
BA42R       & Secondary auditory   cortex                & 1             & 1             & 1               & 1              & 1                & 1                 \\
BA43L       & Postcentral gyrus                          & 1             & 1             & 4               & 1              & 1                & 1                 \\
BA43R       & Postcentral gyrus                          & 1             & 1             & 4               & 1              & 1                & 1                 \\
BA44L       & Opercular part of   inferior frontal gyrus & 1             & 1             & 1               & 3              & 1                & 1                 \\
BA44R       & Opercular part of   inferior frontal gyrus & 1             & 1             & 1               & 3              & NaN              & 1                 \\
BA45L       & Inferior frontal   gyrus                   & 1             & 1             & 1               & 3              & 1                & 1                 \\
BA45R       & Inferior frontal   gyrus                   & 1             & 1             & 1               & 3              & NaN              & 1                 \\
BA46L       & Medial prefrontal   cortex                 & 3             & 4             & 3               & 3              & 3                & 4                 \\
BA46R       & Medial prefrontal   cortex                 & 1             & 4             & 3               & 3              & 3                & NaN               \\
BA47L       & Ventro-lateral   prefrontal cortex         & 4             & 5             & 5               & 3              & 5                & 5                 \\
BA47R       & Ventro-lateral   prefrontal cortex         & 4             & 5             & 5               & 3              & 5                & 5                \\
\hline
\end{longtable}

\vspace{10mm}

\begin{longtable}[H]{ll|cccccc} 
\caption{Modules for High Modularity Group} \\
\hline
\textbf{ } & \textbf{Brodmann Area Name}         & \textbf{Self} & \textbf{Chaplin} & \textbf{Gagaku} & \textbf{Bach} & \textbf{Cronkite} & \textbf{Xhosa} \\
\hline
\endfirsthead
\hline
\textbf{ } & \textbf{Brodmann Area Name}         & \textbf{Self} & \textbf{Chaplin} & \textbf{Gagaku} & \textbf{Bach} & \textbf{Cronkite} & \textbf{Xhosa} \\
\hline
\endhead
BA01L          & Primary somatosensory   cortex                             & 4             & 4             & 4               & 4              & 4                & 4                 \\
BA01R          & Primary somatosensory   cortex                             & 4             & 4             & 4               & 4              & 4                & 4                 \\
BA02L          & Secondary   somatosensory cortex                           & 2             & 4             & 4               & 4              & 4                & 4                 \\
BA02R          & Secondary   somatosensory cortex                           & 4             & 4             & 4               & 4              & 4                & 4                 \\
BA03L          & Tertiary   somatosensory cortex                            & 2             & 4             & 4               & 2              & 4                & 4                 \\
BA03R          & Tertiary   somatosensory cortex                            & 4             & 4             & 4               & 4              & 4                & 4                 \\
BA04L          & Primary motor cortex                  & 2             & 4             & 4               & 4              & 4                & 4                 \\
BA04R          & Primary motor cortex                  & 4             & 4             & 4               & 4              & 4                & 4                 \\
BA05L          & Superior parietal   sulcus            & 2             & 4             & 4               & 2              & 2                & 4                 \\
BA05R          & Superior parietal   sulcus            & 2             & 4             & 4               & 2              & 2                & 4                 \\
BA06L          & Supplementary motor   area            & 3             & 4             & 4               & 4              & 3                & 4                 \\
BA06R          & Supplementary motor   area            & 3             & 4             & 4               & 4              & 3                & 3                 \\
BA07L          & Superior parietal   gyrus             & 2             & 2             & 2               & 2              & 2                & 2                 \\
BA07R          & Superior parietal   gyrus             & 2             & 2             & 2               & 2              & 2                & 2                 \\
BA08L          & Pre-supplementary   motor area                             & 3             & 3             & 3               & 3              & 3                & 3                 \\
BA08R          & Pre-supplementary   motor area                             & 3             & 3             & 5               & 3              & 3                & 3                 \\
BA09L          & Dorsolateral   prefrontal cortex                           & 3             & 3             & 3               & 3              & 3                & 3                 \\
BA09R          & Dorsolateral   prefrontal cortex                           & 3             & 3             & 3               & 3              & 3                & 3                 \\
BA10L          & Fronto-parietal   cortex              & 3             & 3             & 3               & 3              & 3                & 5                 \\
BA10R          & Fronto-parietal   cortex              & 3             & 3             & 3               & 3              & 3                & 5                 \\
BA11L          & Orbital frontal   cortex              & NaN           & NaN           & NaN             & NaN            & NaN              & NaN               \\
BA11R          & Orbital frontal   cortex              & NaN           & NaN           & NaN             & NaN            & NaN              & NaN               \\
BA13L          & Insula                                & 1             & NaN           & 1               & 1              & 1                & 1                 \\
BA13R          & Insula                                & 1             & NaN           & 1               & 1              & 1                & 1                 \\
BA17L          & Primary visual cortex                 & 2             & 2             & 2               & 2              & 2                & 2                 \\
BA17R          & Primary visual cortex                 & 2             & 2             & 2               & 2              & 2                & 2                 \\
BA18L          & Secondary visual   cortex             & 2             & 2             & 2               & 2              & 2                & 2                 \\
BA18R          & Secondary visual   cortex             & 2             & 2             & 2               & 2              & 2                & 2                 \\
BA19L          & Cuneus                                & 2             & 2             & 2               & 2              & 2                & 2                 \\
BA19R          & Cuneus                                & 2             & 2             & 2               & 2              & 2                & 2                 \\
BA20L          & Inferior temporal   gyrus             & 5             & NaN           & 5               & 5              & NaN              & 1                 \\
BA20R          & Inferior temporal   gyrus             & 5             & NaN           & 5               & 5              & NaN              & NaN               \\
BA21L          & Medial temporal gyrus                 & 5             & 1             & 5               & 5              & 1                & 1                 \\
BA21R          & Medial temporal gyrus                 & 3             & 1             & 5               & 5              & 1                & 1                 \\
BA22L          & Superior temporal   gyrus             & 1             & 1             & 1               & 1              & 1                & 1                 \\
BA22R          & Superior temporal   gyrus             & 1             & 1             & 1               & 1              & 1                & 1                 \\
BA23L          & Posterior cingulate   cortex1         & 2             & 2             & 2               & 2              & 2                & 2                 \\
BA23R          & Posterior cingulate   cortex1         & 2             & 2             & 2               & 2              & 2                & 2                 \\
BA24L          & Dorsal anterior   cingulate cortex                         & 3             & 3             & 3               & 4              & 3                & 3                 \\
BA24R          & Dorsal anterior   cingulate cortex                         & 3             & 3             & 3               & 4              & 3                & 3                 \\
BA25L          & Subgenual anterior   cingulate cortex                      & 5             & NaN           & 5               & 5              & NaN              & 5                 \\
BA25R          & Subgenual anterior   cingulate cortex                      & 5             & NaN           & 5               & 5              & NaN              & 5                 \\
BA27L          & Parahippocampal   gyrus1              & NaN           & NaN           & NaN             & NaN            & NaN              & NaN               \\
BA27R          & Parahippocampal   gyrus1              & NaN           & NaN           & NaN             & NaN            & NaN              & NaN               \\
BA28L          & Hippocampal area1                     & 5             & 5             & 5               & 5              & 3                & 5                 \\
BA28R          & Hippocampal area1                     & 5             & 5             & 5               & 5              & NaN              & NaN               \\
BA29L          & Retrosplenial cortex1                 & 2             & 2             & 2               & 2              & 2                & 2                 \\
BA29R          & Retrosplenial cortex1                 & 2             & 2             & 2               & 2              & 2                & 2                 \\
BA30L          & Retrosplenial cortex2                 & 2             & 2             & 2               & 2              & 2                & 2                 \\
BA30R          & Retrosplenial cortex2                 & 2             & 2             & 2               & 2              & 2                & 2                 \\
BA31L          & Posterior cingulate   cortex2         & 2             & 2             & 2               & 2              & 2                & 2                 \\
BA31R          & Posterior cingulate   cortex2         & 2             & 2             & 2               & 2              & 2                & 2                 \\
BA32L          & Pregenual anterior   cingulate cortex                      & 3             & 3             & 3               & 3              & 3                & 3                 \\
BA32R          & Pregenual anterior   cingulate cortex                      & 3             & 3             & 3               & 3              & 3                & 3                 \\
BA33L          & Rostral anterior   cingulate cortex                        & NaN           & NaN           & NaN             & NaN            & NaN              & NaN               \\
BA33R          & Rostral anterior   cingulate cortex                        & NaN           & NaN           & NaN             & NaN            & NaN              & NaN               \\
BA34L          & Hippocampus                           & 5             & 5             & NaN             & 5              & NaN              & 5                 \\
BA34R          & Hippocampus                           & 5             & 5             & 5               & 5              & NaN              & 5                 \\
BA35L          & Hippocampal area2                     & 5             & 5             & 5               & 5              & NaN              & 5                 \\
BA35R          & Hippocampal area2                     & 5             & NaN           & 5               & 5              & NaN              & 5                 \\
BA36L          & Parahippocampal   gyrus2              & 5             & 5             & 5               & 5              & 3                & 5                 \\
BA36R          & Parahippocampal   gyrus2              & 5             & 5             & 5               & 5              & 2                & 5                 \\
BA37L          & Occipital-temporal   cortex           & 2             & 2             & 2               & 2              & 2                & 2                 \\
BA37R          & Occipital-temporal   cortex           & 2             & 2             & 2               & 2              & 2                & 2                 \\
BA38L          & Temporal pole                         & 5             & 5             & 5               & 5              & 1                & 5                 \\
BA38R          & Temporal pole                         & 5             & 5             & 5               & 5              & 3                & 5                 \\
BA39L          & Angular gyrus                         & 2             & 2             & 2               & 2              & 2                & 2                 \\
BA39R          & Angular gyrus                         & 2             & 2             & 2               & 2              & 2                & 2                 \\
BA40L          & Intra-parietal sulcus                 & 2             & 4             & 4               & 4              & 2                & 4                 \\
BA40R          & Intra-parietal sulcus                 & 2             & 4             & 4               & 4              & 2                & 4                 \\
BA41L          & Primary auditory   cortex             & 1             & 1             & 1               & 1              & 1                & 1                 \\
BA41R          & Primary auditory   cortex             & 1             & 1             & 1               & 1              & 1                & 1                 \\
BA42L          & Secondary auditory   cortex           & 1             & 1             & 1               & 1              & 1                & 1                 \\
BA42R          & Secondary auditory   cortex           & 1             & 1             & 1               & 1              & 1                & 1                 \\
BA43L          & Postcentral gyrus                     & 1             & 4             & 1               & 1              & 1                & 1                 \\
BA43R          & Postcentral gyrus                     & 1             & 1             & 1               & 1              & 1                & 1                 \\
BA44L          & Opercular part of   inferior frontal gyrus                 & 1             & NaN           & 1               & 3              & 1                & NaN               \\
BA44R          & Opercular part of   inferior frontal gyrus                 & 1             & NaN           & 1               & 3              & 1                & NaN               \\
BA45L          & Inferior frontal   gyrus              & 3             & 3             & 1               & 3              & 1                & 1                 \\
BA45R          & Inferior frontal   gyrus              & 3             & NaN           & 1               & 3              & NaN              & 1                 \\
BA46L          & Medial prefrontal   cortex            & 3             & 3             & 1               & 3              & NaN              & 3                 \\
BA46R          & Medial prefrontal   cortex            & 3             & 3             & NaN             & 3              & NaN              & 3                 \\
BA47L          & Ventro-lateral   prefrontal cortex                         & 3             & 5             & 5               & 3              & 3                & 5                 \\
BA47R          & Ventro-lateral   prefrontal cortex                         & 3             & 5             & 5               & 3              & 3                & 5               \\
\hline 
\end{longtable}

\bibliography{SupplementalMaterial}